\documentclass{emulateapj}
\usepackage{amssymb}
\usepackage{apjfonts}
\usepackage{graphicx}
\begin{document}
\title{Disappearance of a narrow Mg II absorption system with a measured velocity up to $\rm 166,000~km~s^{-1}$}

\shorttitle{The disappearance of a Mg II absorption system with
relativistic velocity}

\shortauthors{Chen \& Qin}
\author{Zhi-Fu Chen\altaffilmark{1, 2} and Yi-Ping Qin\altaffilmark{1, 2, 3}, }
 \altaffiltext{1}{Department of Physics and Telecommunication Engineering, Baise
University, Baise 533000, China; zhichenfu@126.com}
 \altaffiltext{2}{Center for Astrophysics, Guangzhou University, Guangzhou 510006, China}
\altaffiltext{3}{Physics Department, Guangxi University, Nanning
530004, China }

\begin{abstract}
Quasar J152645.61+193006.7 ($z_{\rm e}=2.5771$) was observed by the
Sloan Digital Sky Survey (SDSS) on 2006 May 31, and again on 2011
April 9. The time interval of the SDSS two observations is 497.4
days at the quasar rest frame. From the spectra of this quasar, we
detect a phenomenon of disappearance of a narrow $\rm
Mg~II~\lambda\lambda2796,2803$ absorption system with a velocity up
to $\rm 166,129~km~s^{-1}$ with respect to the quasar. This
disappearance event can be caused by changes in the ionization state
of absorbing gas or by the bulk motion of the absorbing gas across
the quasar sightline. The coverage fraction analysis shows that this
absorber partially covers the background emission sources with an
effective coverage fraction of $C_{\rm f}=0.40\pm0.06$. The time
variation analysis and the coverage fraction analysis imply that
this absorber might be intrinsic to the quasar. However, the
scenario of a cosmologically separated foreground object located at
$z=0.9170$ accounting for the phenomenon cannot be ruled out
according to current available data.
\end{abstract}

\section{Introduction}
The outflow of quasar is often believed to be the gas blown away
from the accretion disk. Later, the gas can be accelerated by the
radiation pressure (Castor et al. 1975; Proga et al. 1998;
Dorodnitsyn\& Novikov 2005; Proga \& Kurosawa 2009, and references
therein), the magnetic force (Contopoulos \& Lovelace 1994; Li 1996;
Kudoh \& Shibata 1997a,b; Lery et al. 1998,1999; Proga 2007, and
references therein) and the thermal driving (Begelman et al. 1983;
Kallman 2005; Proga 2007; Owen et al. 2012, and references therein).
If the gas is heated to a high temperature reaching up to the
Compton temperature, the thermal driving would be effective enough
to produce a thermal wind (e.g., Proga \& Kallman 2002; Chelouche \&
Netzer 2005). The thermal driving might be less important if the
temperature of gas is well bellow the Compton temperature (Proga
2007).

The outflow was detected most conspicuously via blueshifted
absorption lines. These lines can be split into three categories
based on line widths: broad absorption lines (BALs), with the
typical line widths being broader than $2000\rm~km~s^{-1}$ at depths
$>10\%$ below the continuum (Weymann et al. 1991); narrow absorption
lines (NALs), with absorption troughs being narrower than a few
hundred $\rm~km~s^{-1}$; and intermediate mini-broad absorption
lines (mini-BALs), with line widths lying between those of BALs and
NALs. They can appear at a very wide range of speeds from
$10^2\rm~km~s^{-1}$ to $10^5\rm~km~s^{-1}$ in multiband spectra
(e.g., X-ray, UV and optical spectra) (e.g., Misawa et al. 2007;
Ganguly \& Brotherton 2008; Tombesi et al. 2010, 2011). Absorption
lines of highly ionized elements (e.g., $\rm O~VIII, Fe~XVIII$) have
been observed only in high energy band spectra (e.g., X-ray
spectra). Nevertheless, transitions by the $\rm C~IV$, $\rm Si~IV$,
$\rm N~V$, and $\rm O~VI$ can be seen in UV but not in X-ray spectra
(e.g., Kaspi et al. 2002; Krongold et al. 2003).

Blueshifted absorption lines are very common in quasar spectra.
However, only a small number of them are intrinsic to quasars. Most
of them are intervening absorption lines, which are believed to be
physically associated with the cosmologically intervening foreground
galaxies lying on quasar sightlines (Bergeron 1986). BALs are
undoubtedly intrinsic to the quasar (Murray et al. 1995; Elvits
2000). Both intrinsic and intervening NALs usually show similar line
widths and line strengths, thereby it is difficult to determine
which NALs are truly intrinsic to the quasars (Qin et al. 2013). One
can separate intrinsic NALs from intervening NALs via several
criteria: time variation, partial coverage, line profile, high
ionization state, and so on (e.g., Barlow \& Sargent 1997; Hamann et
al. 1997, 2011; Misawa et al. 2005; Chen et al. 2013a). Among these
methods, the time variation analysis and the coverage fraction
analysis are the most effective and most frequently utilized ones
(Ganguly et al. 1999; Misawa et al. 2003; Wise et al. 2004; Hamann
et al. 2011).

Variations of intrinsic absorption lines seem to be common, but
extreme cases such as the disappearance and emergence of absorption
lines from the spectra are rare. So far, only a small number of
disappearance (e.g., Hall et al. 2011; Filiz Ak et al. 2012;
Capellupo et al. 2011, 2012, 2013; Chen et al. 2013a) and emergence
(Ma 2002; Hamann et al. 2008; Leighly et al. 2009; Krongold et al.
2010; Rodr\'Iguez Hidalgo et al. 2011; Vivek et al. 2012; Chen et
al. 2013b) cases are reported. Such variations, in principle, can be
caused by changes in the ionization state of absorbing gas and the
coverage fraction of the absorber to the background emission
sources. The change in the coverage fraction could be driven via the
motions of absorbing gas (e.g., Proga \& Kallman 2002; Leighly et
al. 2009; Chen et al. 2013a). The change in ionization state of
absorbing gas could arise from the variations of background
emissions (e.g., Hamann et al. 2011).

Absorption lines with very high ionization states (such as $\rm
Fe~XXV$) can arise from the outflow with a speed up to $\rm
10^{5}~km~s^{-1}$ (e.g., Tombesi et al. 2011), while lines
with lower ionization states (such as $\rm
C~IV\lambda\lambda1548,1551$) in the outflow can be searched
effectively at a velocity of up to $\rm \sim 70,000~km~s^{-1}$
without blending with $\rm Ly\alpha$ forest. (Misawa et al. 2007).
Chen et al. (2013a) firstly reported the disappearance of a narrow
$\rm Mg~II\lambda\lambda2796,2803$ absorption doublet to be formed
in an outflow with a speed of $\rm 8,423~km~s^{-1}$. However, no
disappearance events of narrow absorption lines with speeds up to
$\rm 10^5~km~s^{-1}$ relative to the quasar system have ever been
detected. In this paper, we report the disappearance of a narrow
$\rm Mg~II\lambda\lambda2796,2803$ absorption system with a measured
speed up to $\rm 166,129~km~s^{-1}$ relative to the quasar emission
redshift, discovered from the spectrum of SDSS J152645.61+193006.7.

Throughout this paper the cosmological parameters of $\rm
\Omega_\Lambda=0.7$, $\rm \Omega_M=0.3$ and $\rm H_{0}= 70 ~\rm km~
s^{-1}~ Mpc^{-1}$ are adopted.

\section{Spectral analysis}
Up to $7,932$ quasars contained in the quasar spectroscopic catalog
of Sloan Digital Sky Survey Data Release Seven (SDSS-I/II, York et
al. 2000; Schneider et al. 2010) were re-observed by SDSS-III
(Eisenstein et al. 2011; Ross et al. 2012; P\^aris et al. 2012),
providing us a chance to measure possible disappearance of
absorption systems. NALs are often confused by the significant
systematic sky-subtraction residuals longward of $7000$ \AA~ in
SDSS-I/II spectra (Wild \& Hewett 2005). Our analysis focuses on
narrow $\rm Mg~II\lambda\lambda2796,2803$ absorption systems with a
very large value of measured velocities relative to the
corresponding quasars. We conservatively constrain our analysis
within the wavelength range shortward of $7000$ \AA~ at the
observed-frame and longward of the $\rm Ly\alpha$ emission line, and
pay our attention to the absorption of the doublets with blueshifted
velocities $\rm>50,000~km~s^{-1}$ with respect to the quasar
emission redshift, and select only those quasars with their median
SNR over the whole spectrum being greater than 5 (the median SNR
over the whole spectrum is defined by P\^aris et al. 2012). Spectra
of $4,276$ quasars satisfy these criteria. Adopting the method of
the combination of cubic splines (for underlying continuum) and
Gaussians (for emission line features), we obtain a pseudo-continuum
for each spectrum of these quasars by fit (Nestor et al. 2005; Chen
et al. 2013a). This pseudo-continuum is used to normalize spectral
fluxes and flux uncertainties. We search narrow $\rm
Mg~II~\lambda\lambda2796,2803$ absorption doublets in the
pseudo-continuum normalized spectrum. The rest-frame equivalent
widths ($W_{\rm r}$) of the detected absorption lines are measured
from Gaussian fittings. The corresponding uncertainties are
estimated via
\begin{equation}
(1+z)\sigma_w=\frac{\sqrt{\sum_i
P^2(\lambda_i-\lambda_0)\sigma^2_{f_i}}}{\sum_i
P^2(\lambda_i-\lambda_0)}\Delta\lambda,
\end{equation}
where $P(\lambda_i-\lambda_0)$ is the line profile centered at
$\lambda_0$, $\lambda_i$ is the wavelength, and $\sigma_{f_i}$ is
the normalized flux uncertainty as a function of pixel (Nestor et
al. 2005; Chen et al. 2013a). The sum is performed over an integer
number of pixels that covers at least $\pm 3$ characteristic
Gaussian widths.

We notice a single phenomenon of disappearance of narrow $\rm
Mg~II~\lambda\lambda2796,2803$ absorption doublets with velocities
up to $\rm 10^5~km~s^{-1}$. For quasar SDSS J152645.61+193006.7
($\rm z_e=2.5771$, taken from Hewett \& Wild 2010), we observe that
a narrow $\rm Mg~II~\lambda\lambda2796,2803$ absorption system with
$z_{\rm abs} =0.9170$ ($FWHM\rm_{\lambda2796}=160~km~s^{-1}$), which
corresponds to $\rm 166,129~km~s^{-1}$ with respect to the quasar,
imprinted in the SDSS-I/II spectrum disappeared from the SDSS-III
spectrum. Both spectra, with the pseudo-continuum fittings, of the
quasar are displayed in Fig. 1. The corresponding pseudo-continuum
normalized spectra are presented in Fig. 2. The measurements of the
corresponding absorption lines are presented in Table 1 (see also
Fig. 2).
\begin{figure*}
\centering
\includegraphics[width=17 cm,height=6.8 cm]{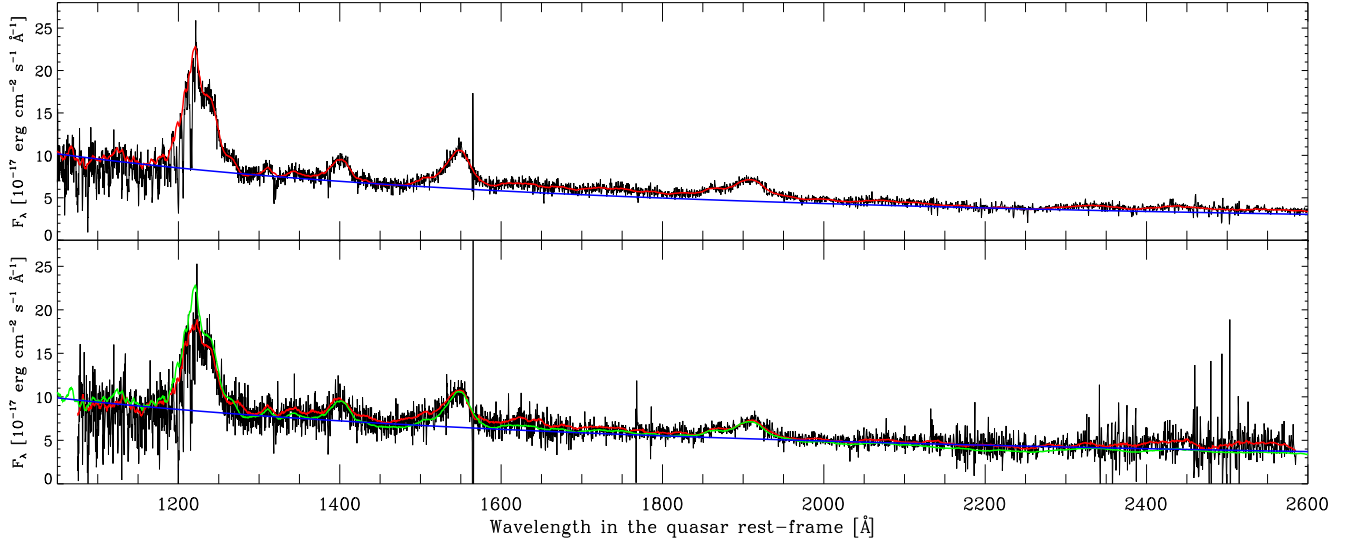}
\vspace{2ex} \caption{The spectra of quasar J152645.61+193006.7 with
$z_e=2.5771$, observed by SDSS-I/II (the lower panel) and by
SDSS-III (the upper panel), respectively. The blue solid lines
represent the power-law continuum fittings. The red solid lines
represent the pseudo-continua, and the green solid line presented in
the lower panel is just the pseudo-continuum shown in the upper
panel.}
\end{figure*}
\begin{figure*}
\centering
\includegraphics[width=17 cm,height=6.8 cm]{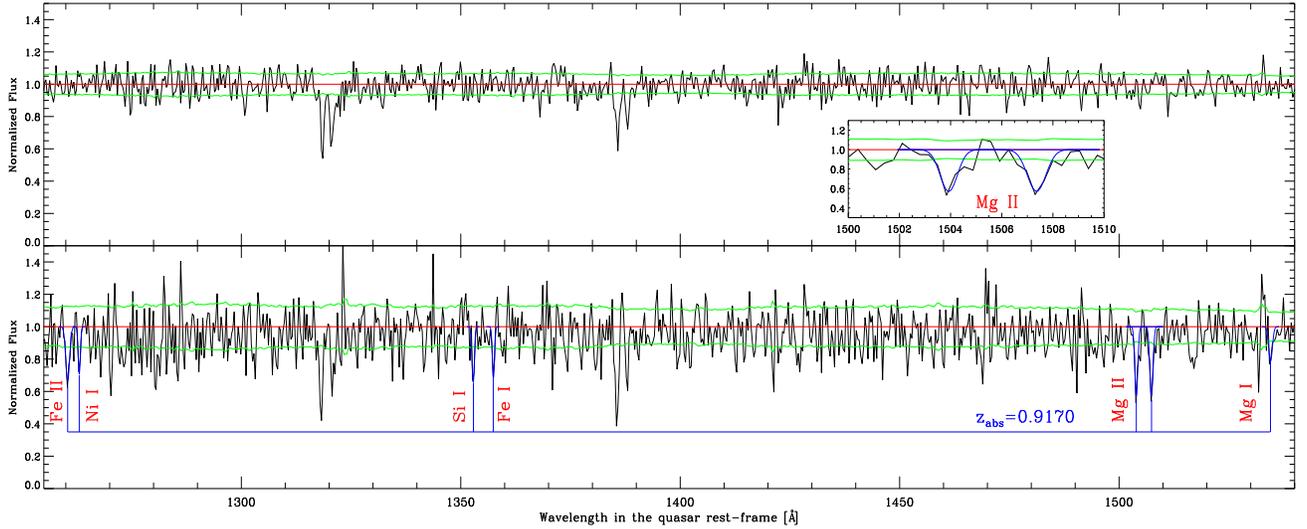}
\vspace{2.ex}\caption{The pseudo-continuum normalized spectra of
quasar J152645.61+193006.7, observed by SDSS-I/II (the lower panel)
and by SDSS-III (the upper panel), respectively. The green cures
represent the flux uncertainty levels which have been normalized by
the corresponding pseudo-continuum, and the blue curves represent
the Gaussian fittings. One narrow $\rm Mg~II\lambda\lambda2796,2803$
absorption system with $z_{\rm abs}=0.9170$ presented in the
SDSS-I/II spectrum disappeared from the SDSS-III spectrum.}
\end{figure*}
\begin{table*}
\centering\caption{Parameters of the disappearing $\rm Mg~II$
absorption system} \tabcolsep 1.7mm
 \begin{tabular}{ccccccccccccccc}
 \hline\hline \scriptsize
&$\rm Mg~II\lambda2796$&$\rm Mg~II\lambda2803$&$\rm Mg~I\lambda2853$&$\rm Fe~I\lambda2523$&$\rm Si~I\lambda2515$&$\rm Ni~I\lambda2348$&$\rm Fe~II\lambda2344$\\
\hline\scriptsize
$W_r$ (\AA)&0.70$\pm$0.15&0.73$\pm$0.16&0.42$\pm$0.18&0.37$\pm$0.14&0.45$\pm$0.14&0.35$\pm$0.13&0.69$\pm$0.23\\
\hline \scriptsize
$W_r^a$ (\AA)&0.09&0.09&0.09&0.07&0.07&0.07&0.11\\
\hline
\end{tabular}
\\
$^a$The equivalent width limits estimated from the SDSS-III spectrum
are calculated by Equation (1) as well.
\end{table*}

\section{Results and Discussion}
Quasar J152645.61+193006.7 was observed by SDSS-I/II on 2006 May 31,
and re-observed by SDSS-III on 2011 April 9. The time interval of
the two observations is 497.4 days at the quasar rest-frame. From
the two SDSS spectra of this quasar, we detect the disappearance of
a narrow $\rm Mg~II~\lambda\lambda2796,2803$ absorption system with
$z_{\rm abs}=0.9170$. This absorption system has a velocity offset
$\rm 166,129~km~s^{-1}$ with respect to the quasar emission redshit
($\rm z_e=2.5771$). Rough estimation from the curve of growth gives
rise to the ionic column density of $N\rm_{\lambda2796} \approx
10^{13.4}~cm^{-2}$.

\subsection{Coverage fractions} The optical depth ratio of
the $\rm Mg~II~\lambda\lambda2796,2803$ absorption doublet has a
value of $\rm \tau_{2796}:\tau_{2803}\approx2:1$, expected from the
atomic physics (Savage \& Sembach 1991; Verner et al. 1994). If the
absorber incompletely covers the background emission sources, the
photons apparently leaking through the absorption line region will
give rise to a value of optical depth ratio deviated from the
theoretical value (e.g., Barlow \& Sargent 1997; Hamann et al. 1997;
Crenshaw et al. 1999). In principle, the intrinsic absorber is often
expected to partially cover the background emission sources. In
order to check this, we fit the weaker member of the $\rm
Mg~II\lambda\lambda2796,2803$ doublet with a Gaussian profile, scale
the model in terms of what atomic physics expects for the stronger
member, and compare it with the data. The results are illustrated in
Fig. 3. It can be clearly seen from the figure that the profile of
$\rm \lambda2796$ scaled to the atomic physics is inconsistent with
the data, which is indicative of partial coverage.
\begin{figure}
\centering
\includegraphics[width=6.5 cm,height=4.9 cm]{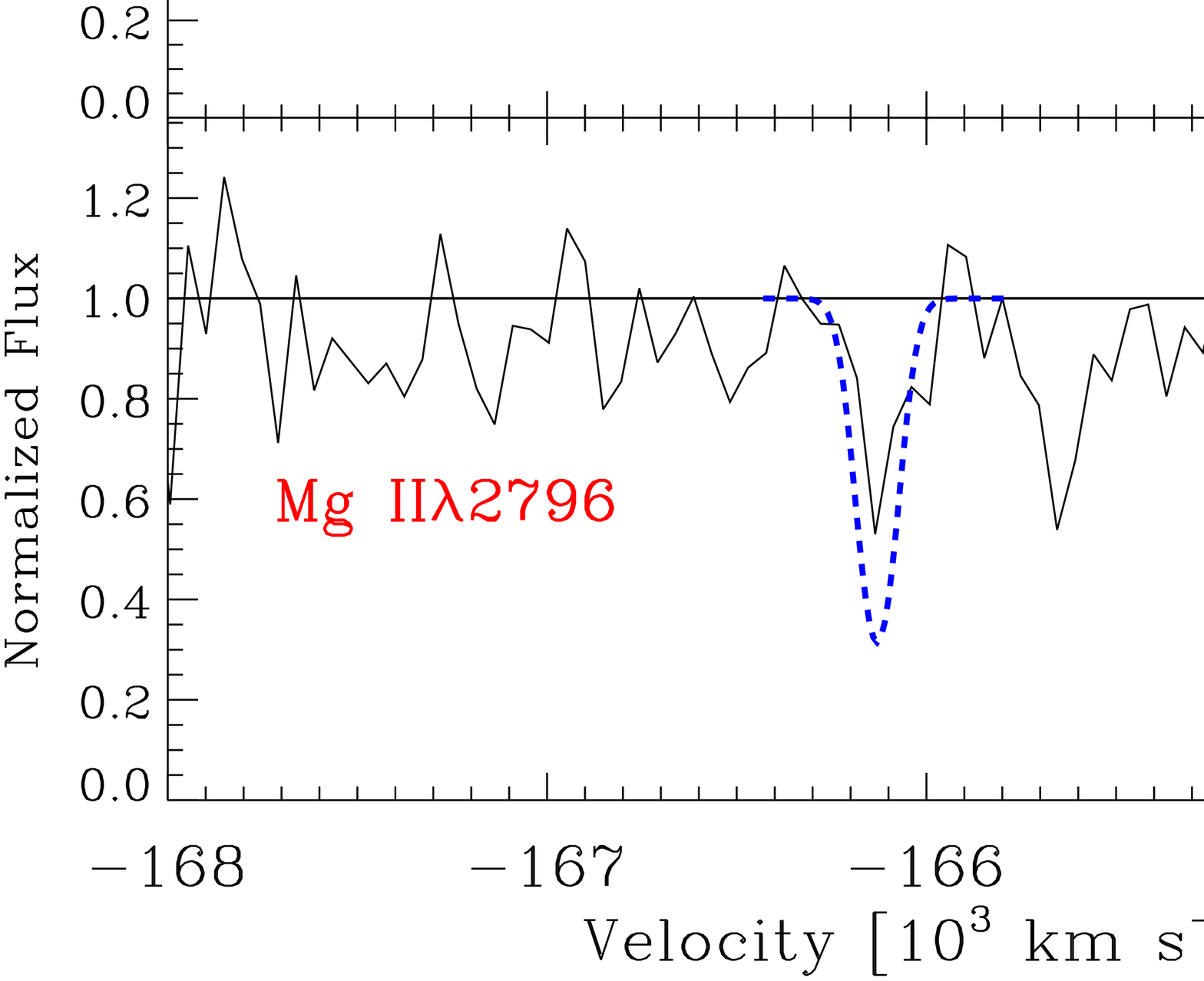}
\vspace{6ex}\caption{The $\rm Mg~II\lambda\lambda2796,2803$
absorption with a velocity of $166,129~\rm km~s^{-1}$ relative to
the quasar emission redshift. The Gaussian fitting to the weaker
member ($\lambda2803$) is shown in the upper panel with blue
short-dash lines. The profile of the stronger member ($\lambda2796$)
that the atomic physics predicts is shown in the lower panel with
blue short-dash lines.}
\end{figure}

The effective coverage fraction of the absorber to the background
emission sources can be computed from the residual intensities of
the resonance doublet. The normalized residual intensity as a
function of velocity from the line center is
\begin{equation}
R(v)=[1-C_{\rm f}(v)]+C_{\rm f}(v)e^{\rm -\tau(v)}
\end{equation}
where $C_{\rm f}(v)$ is the effective fraction and $\tau(v)$ is the
optical depth at velocity $v$. For the $\rm
Mg~II~\lambda\lambda2796,2803$ doublet, the effective coverage
fraction can be evaluated via
\begin{equation}
C_{\rm f}(v)=\frac{[R_{\rm r}(v)-1]^2}{R_{\rm b}(v)-2R_{\rm r}(v)+1}
\end{equation}
where the subscript $r$ and $b$ refer to the redder and bluer
members of the $\rm Mg~II~\lambda\lambda2796,2803$ doublet (e.g.,
Hamann et al. 1997; Barlow \& Sargent 1997; Crenshaw et al. 1999;
Misawa et al. 2005, 2007). Due to the low resolution of SDSS
spectra, it is inappropriate to evaluate the effective coverage
fraction pixel by pixel for the narrow $\rm
Mg~II~\lambda\lambda2796,2803$ absorption doublet. Therefore, in
this paper, we use the normalized residual intensities at line cores
to evaluate the effective coverage fraction of the $\rm
Mg~II~\lambda\lambda2796,2803$ absorber, and get $C_{\rm
f}=0.40\pm0.06$.

In principle, the intrinsic absorber could harbor different coverage
fractions to the continuum source and the emission line regions
(e.g., Ganguly et al. 1999; Gabel et al. 2005). Therefore, the
effective coverage fraction is the fraction of the photons from all
the background emission sources going through the intrinsic
absorber. Here, we consider the situation that the background
photons only arise from the continuum source and the broad emission
line region (BELR) (e.g., Ganguly et al. 1999; Wu et al. 2010; Chen
et al. 2013b), and assume that the optical depths are the same from
the two emission sources. In this case, the effective coverage
fraction is the weighted average of the coverage fractions of the
two regions. That is
\begin{equation}
C_{\rm f}=\frac{C_{\rm c}+WC_{\rm e}}{1+W},
\end{equation}
where $C_{\rm e}$ and $C_{\rm c}$ are the coverage fractions of the
BELR and the continuum source, and $W=f_{\rm e}/f_{\rm c}$ is the
flux ratio of the broad emission line (without continuum) and the
continuum at the wavelength of the absorption line (e.g., Ganguly et
al. 1999; Misawa et al. 2005, 2007; Wu et al. 2010). In order to
evaluate the value of $W$, we use the method adopted by Chen et al.
(2009) to fit a power-law continuum ($f\propto\nu^{-\alpha}$) for
the quasar spectra of J152645.61+193006.7, where, several spectral
regions without obvious emission lines are selected. We obtain
$\alpha=-1.34$ from the SDSS-III spectrum and $\alpha=-1.08$ from
the SDSS-I/II spectrum. (These continua are also plotted in Fig. 1
with blue curves.) The strengths, which are measured from the $\rm
C~IV$ broad emission line and the power-law continuum at the
wavelength of the $\rm Mg~II$ absorption doublet, give rise to a
value of $W=0.23$. Both values of $C_{\rm c}$ and $C_{\rm e}$ can
not be determined independently. However, one can constrain a
relation for them, which is plotted in Fig. 4.
\begin{figure}
\centering
\includegraphics[width=6.5 cm,height=4.9 cm]{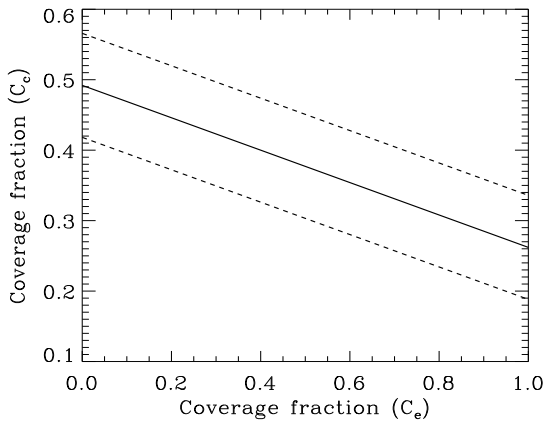}
\vspace{6ex}\caption{The $C_{\rm c}$ --- $C_{\rm e}$ parameter
plane, showing solutions allowed by equation (4). The solid line is
determined by equation (5) with a nominal value of $C_{\rm f}=0.4$,
and the dash lines correspond to $1\sigma$ errors of $C_{\rm f}$.}
\end{figure}

From equation (4), one can obtain the expression of $C_{\rm c}$ as a
function of $C_{\rm f}$, $C_{\rm e}$ and $W$, which is
\begin{equation}
C_c=(1+W)C_f+WC_e=1.23C_f-0.23C_e
\end{equation}
where $W=0.23$ and $C_{\rm f}=0.40\pm0.06$. Although the value of
$C_{\rm c}$ can not be determined without knowing $C_{\rm e}$, we
can derive the upper and lower limits of $C_{\rm c}$ from this
equation by considering two extreme values of $C_{\rm e}$: $C_{\rm
e}=0$ and $C_{\rm e}=1$. They correspond to $C_{\rm c}=0.49$ and
$C_{\rm c}=0.26$, respectively.

\subsection{The origin of time variation}
Time variation of absorption lines and partial coverage of the
absorber to background emission sources are the two most popular
indicators to separate the intrinsic NALs from the intervening NALs.
The coverage fraction and time variation analysis (refer to Fig.
3and Table 1) imply that the narrow $\rm
Mg~II~\lambda\lambda2796,2803$ absorption system with $z_{\rm
abs}=0.9170$ might be truly intrinsic to the quasar
J152645.61+193006.7 with $\rm z_e=2.5771$. If so, what would be the
mechanism driving the disappearance of this $\rm
Mg~II~\lambda\lambda2796,2803$ absorption system?

Time variations of absorption lines can be induced by changes in the
ionization state of the absorbing gas, which can be driven by the
variations of background emissions (from the continuum source or/and
the BELR) or caused by a screen of variable optical depth between
the absorber and the continuum source (e.g., Misawa et al. 2007), or
by the bulk motion of the absorbing gas across our sightline (that
might give rise to a change in the coverage fraction of the absorber
to background emission sources). If the disappearance of the $\rm
Mg~II~\lambda\lambda2796,2803$ absorption system is due to the
fluctuation of background emissions, it is expectable that an
observable change of quasar ionized radiations should be noticed
during the corresponding period of observations, as the
disappearance event is an extreme case of line variation. In fact,
it can be clearly seen from Fig. 1 (refer to the green and red solid
curves shown in the lower panel) that the quasar emissions at the
two epochs of observation are quite stable. This suggests that the
disappearance is unlikely to be caused by the fluctuation of the
quasar emissions.

A screen of variable optical depth between the absorber and the
background emission sources can also bring about a fluctuation of
the incidence flux of the absorber (e.g., Misawa et al. 2007), which
is capable of causing a change in the absorber's ionization levels.
If so, the variable ionization scenario may still be alive even if
the quasar luminosity is stable. Unfortunately, conditions in the
screen can not be determined with present data.

Perhaps, on the rest-frame time corresponding to 2006 May 31 (when
quasar J152645.61+193006.7 was observed by SDSS-I/II), the sightline
of the object possibly passed through the absorber (possibly near
its edge), while on that corresponding to 2011 April 9 (when the
object was observed by SDSS-III), the absorber had already moved
away from the sightline due to its proper motion. If so, the
disappearance of absorption troughs would be expectable. In this
way, a notable change of quasar emissions is not required.
Therefore, it is likely that the disappearance event of the $\rm
Mg~II~\lambda\lambda2796,2803$ absorption system is due to the
proper motion of the absorption gas across our sightline. Most of
the UV continuum emission originates from the inner region of a
geometric thin and optical thick accretion disk, whose size scale is
$D_{\rm cont} \sim 5R_{\rm S}=10GM_{\rm BH}/c^2$ (Wise et al. 2004;
Misawa et al. 2005). Let us consider the case of the virial black
hole mass estimated by Shen et al. (2011) as the black hole mass,
$M_{\rm BH}$. That is $M_{\rm BH}=10\rm^{9.51}M_\odot$. We then
obtain a value of $D_{\rm cont}$, $4.8\times10\rm^{15}~cm$. In terms
of the coverage fraction of the absorber to the continuum emission
source, namely from $26\%$ to $49\%$, the lower limits of the
absorber radius are from $1.2\times10\rm^{15}~cm$ to
$2.4\times10\rm^{15}~cm$. Assuming a face-on accretion disk and that
the movement of the absorber is perpendicular to our sightline, one
can derive the transverse velocity of the absorber by $v_{\rm
cross}=D_{\rm cont}/t_{\rm tr}$, where $t_{\rm tr}$ is the time
interval of the two SDSS observations, which is $497.4$ days. That
yields $v_{\rm cross}=256\rm~km~s^{-1} \sim 545\rm~km~s^{-1}$. Since
the time interval of $497.4$ days is just the upper limit of transit
time of the absorber, the derived values of $v_{\rm cross}$ are only
the lower limits of the transverse velocity of the absorber.

It is possible as well that a cosmologically separated foreground
galaxy lied on the quasar sightline and located at $z=0.9170$
brought about absorptions to the quasar light, which can be formed
in the halo, star-burst wind, or inflow of the galaxy. The $\rm
Mg~II$ absorption system contains some neutral species such as $\rm
Mg~I~\lambda2853$ and $\rm Fe~I~\lambda2523$, which means that the
absorber should be in very low ionization condition and that its
size can be quite small (in subparsec or smaller) with very high
volume density (e.g., Jones et al. 2010). The crossing velocity
($v_{\rm cross} = 256 - 545~\rm km~s^{-1}$) could also be explained
by rotational velocities of typical spiral galaxies and/or velocity
dispersions of cluster of galaxies. If there exist time variations
of the galaxy emission, a change in the ionization state of the
absorbing gas would be expected, and then it might give rise to the
disappearance of absorption troughs. However, with present data, we
can not tell if the disappearance of $Mg~II~\lambda\lambda2796,2803$
absorption system is truely associated with a cosmologically
separated foreground galaxy.

\section{Summary}
Quasar J152645.61+193006.7 ($z_{\rm e}=2.5771$) was first observed
by the SDSS-I/II on 2006 May 31, and re-observed by the SDSS-III on
2011 April 9, spending a time interval of 497.4 days. We identify
one narrow $\rm Mg~II~\lambda\lambda2796,2803$ absorption system
with $z_{\rm abs}=0.9170$ from the SDSS-I/II spectrum, which has a
relative velocity of $ v_{\rm r}=166,129~\rm km~s^{-1}$ with respect
to the quasar emission redshift. However, this $\rm
Mg~II~\lambda\lambda2796,2803$ absorption system can not be detected
from the SDSS-III spectrum. The coverage fraction analysis shows
that this absorber partially covers the background emission sources
with an effective coverage fraction of $C_{\rm f}=0.40\pm0.06$.

Time variations of the $\rm Mg~II~\lambda\lambda2796,2803$
absorption system might be caused by the change in the ionization
state of absorbing gas or by the motion of the absorber
perpendicular to the quasar sightline. The quasar emissions (from
the continuum source and the broad emission line region) are stable
for the two SDSS observations, suggesting that the changes in the
absorber's ionization condition are unlikely to be caused by the
variation of the quasar emissions. Therefore, if there exist changes
in the ionization state of absorbing gas, they might be driven by a
screen of variable optical depth between the absorber and the
background emissions (Missawa et al. 2007), if the absorber is
intrinsic to the quasar.

The line variation and the partial coverage are the two popular
indicators to tell whether the absorber is truly intrinsic to the
quasar. Accordingly, the $\rm Mg~II~\lambda\lambda2796,2803$
absorption system analyzed in this work is likely to be intrinsic to
the corresponding quasar. However, with present data, we can not
rule out the possibility that a cosmologically separated foreground
galaxy located at $z=0.9170$ gives rise to this absorption system,
since time variations of the galaxy emission and motions of the
cosmologically intervening absorbers might also exist and that would
be capable of causing dramatic variations of absorption lines as
well.

\acknowledgements We thank the anonymous referee for helpful
comments and suggestions. This work was supported by the National
Natural Science Foundation of China (NO. 11363001; No. 11073007),
the Guangxi Natural Science Foundation (2012jjAA10090),  the
Guangzhou technological project (No. 11C62010685), and the Guangxi
university of science and technology research projects (NO.
2013LX155).


\begin{thebibliography}{99}
\bibitem[\protect\citeauthoryear{}{}]{}Begelman, M. C., McKee, C. F., \& Shields, G. A. 1983, ApJ, 271, 70
\bibitem[\protect\citeauthoryear{}{}]{}Bergeron, J. 1986, A\&A, 155, L8
\bibitem[\protect\citeauthoryear{}{}]{}Barlow, T. A., \& Sargent, W. L. W. 1997, AJ, 113, 136
\bibitem[\protect\citeauthoryear{}{}]{}Castor, J. I., Abbott, D. C., \& Klein, R. I. 1975, ApJ, 195, 157
\bibitem[\protect\citeauthoryear{}{}]{}Contopoulos, J., \& Lovelace, R. V. E. 1994, ApJ, 429, 139
\bibitem[\protect\citeauthoryear{}{}]{}Crenshaw, D. M., Kraemer, S. B., Boggess, A., et al.. 1999, ApJ, 516, 750
\bibitem[\protect\citeauthoryear{}{}]{}Chelouche, D., \& Netzer, H. 2005, ApJ, 625, 95
\bibitem[\protect\citeauthoryear{}{}]{}Chen, Z. Y., Gu, M. F., \& Cao, X. W., 2009, MNRAS, 397, 1713
\bibitem[\protect\citeauthoryear{}{}]{}Capellupo, D. M., Hamann, F., Shields, J. C., Rodr\'Iguez Hidalgo, P., \& Barlow, T. A. 2011, MNRAS, 413, 908
\bibitem[\protect\citeauthoryear{}{}]{}Capellupo, D. M., Hamann, F., Shields, J. C., Rodr\'Iguez Hidalgo, P., \& Barlow, T. A. 2012, MNRAS, 422, 3249
\bibitem[\protect\citeauthoryear{}{}]{}Capellupo, D. M., Hamann, F., Shields, J. C., Halpern, J. P., \& Barlow, T. A. 2013, MNRAS, 429, 1872
\bibitem[\protect\citeauthoryear{}{}]{}Chen, Z. F., Chen, Z. Y., Qin, Y. P., et al. 2011, RAA, 4, 401
\bibitem[\protect\citeauthoryear{}{}]{}Chen, Z. F., Qin, Y. P., \& Gu, M. F. 2013a, ApJ, 770, 59
\bibitem[\protect\citeauthoryear{}{}]{}Chen, Z. F., Li, M. S., Huang, W. R., Pan, C. J., \& Li, Y. B. 2013b, doi:10.1093/mnras/stt1247
\bibitem[\protect\citeauthoryear{}{}]{}Dorodnitsyn, A. V. \& Novikov, I. D. 2005, ApJ, 621, 932
\bibitem[\protect\citeauthoryear{}{}]{}Elvits, M. 2000, ApJ, 545, 63
\bibitem[\protect\citeauthoryear{}{}]{}Eisenstein, D. J., Weinberg, D. H., Agol, E., et al. 2011, AJ, 142, 72
\bibitem[\protect\citeauthoryear{}{}]{}Filiz Ak, N., Brandt, W. N., Hall, P. B., et al. 2012, ApJ, 757, 114
\bibitem[\protect\citeauthoryear{}{}]{}Ganguly, R., Eracleous, M., Charlton, J. C., \& Churchill, C. W. 1999, AJ, 117, 2594
\bibitem[\protect\citeauthoryear{}{}]{}Ganguly, R., \& Brotherton, M. S. 2008, ApJ, 672, 102
\bibitem[\protect\citeauthoryear{}{}]{}Gabel, J., Kraemer, S. B., Crensha, D. M., et al. 2005, ApJ, 623, 85
\bibitem[\protect\citeauthoryear{}{}]{}Hamann, F., Barlow, T. A., Junkkarinen, V., \& Burbidge, E. M. 1997, ApJ, 478, 80
\bibitem[\protect\citeauthoryear{}{}]{}Hamann, F., Kaplan, K. F., Rodr\'Iguez Hidalgo, P., Prochaska, J. X., \& Herbert-Fort, S. 2008, MNRAS, 391, L39
\bibitem[\protect\citeauthoryear{}{}]{}Hamann, F., Kanekar, N., Prochaska, J. C., Murphy, M. T., et al. 2011, MNRAS, 410, 1957
\bibitem[\protect\citeauthoryear{}{}]{}Hewett, P. C., \& Wild, V. 2010, MNRAS, 405, 2302
\bibitem[\protect\citeauthoryear{}{}]{}Hall, P. B., Anosov, K., White, R. L., et al. 2011, MNRAS, 411, 2653
\bibitem[\protect\citeauthoryear{}{}]{}Jones, T. M., Misawa, T., Charlton, J. C., Mshar, A. C., \& Ferland, G. J. 2010, ApJ, 715, 1497
\bibitem[\protect\citeauthoryear{}{}]{}Kudoh, T., Shibata, K. 1997a, ApJ, 474, 362
\bibitem[\protect\citeauthoryear{}{}]{}Kudoh, T., Shibata, K. 1997b, ApJ, 476, 632
\bibitem[\protect\citeauthoryear{}{}]{}Kaspi, S., et al. 2002, ApJ, 574, 643
\bibitem[\protect\citeauthoryear{}{}]{}Krongold, Y., Nicastro, F., Brickhouse, N. S., et al. 2003, ApJ, 597, 832
\bibitem[\protect\citeauthoryear{}{}]{}Krongold, Y., Nicastro, F., Elvis, M., et al. 2005, ApJ, 620, 165
\bibitem[\protect\citeauthoryear{}{}]{}Krongold, Y., Binette, L., \& Hernandez-Ibarra, F. 2010, ApJ, 724, 203
\bibitem[\protect\citeauthoryear{}{}]{}Kallman, T. 2005, ASPC, 337, 169
\bibitem[\protect\citeauthoryear{}{}]{}Li, Z. Y. 1996, ApJ, 465, 855
\bibitem[\protect\citeauthoryear{}{}]{}Lery T., Heyvaerts J., Appl S., \& Norman C. A., 1998, A\&A 337, 603
\bibitem[\protect\citeauthoryear{}{}]{}Lery T., Heyvaerts J., Appl S., \& Norman C. A., 1999, A\&A 347, 1055
\bibitem[\protect\citeauthoryear{}{}]{}Leighly, K. M., Hamann, F., Casebeer, D. A., \& Grupe, D. 2009, ApJ, 701, 176
\bibitem[\protect\citeauthoryear{}{}]{}Murray, N., Chiang, J., Grossman, S. A., \& Voit, G. M. 1995, ApJ, 451, 498
\bibitem[\protect\citeauthoryear{}{}]{}Ma, F. 2002, MNRAS, 335, L99
\bibitem[\protect\citeauthoryear{}{}]{}Misawa, T., Yamada, T., Takada-Hidai, M., et al. 2003, AJ, 125, 1336
\bibitem[\protect\citeauthoryear{}{}]{}Misawa, T., Eracleous, M., Charlton, J. C., \& Tajitsu, A. 2005, ApJ, 629, 115
\bibitem[\protect\citeauthoryear{}{}]{}Misawa, T., Charlton, J. C., Eracleous, M., et al. 2007, ApJS, 171, 1
\bibitem[\protect\citeauthoryear{}{}]{}Misawa, T., Eracleous, M., Charlton, J. C., \& Kashikawa, N. 2007, ApJ, 660, 152
\bibitem[\protect\citeauthoryear{}{}]{}Nestor, D. B., Turnshek, D. A., \& Rao, S. M. 2005, ApJ, 628, 637
\bibitem[\protect\citeauthoryear{}{}]{}Owen, J. E., Clarke, C. J., \& Ercolano, B., 2012, MNRAS, 422, 1880
\bibitem[\protect\citeauthoryear{}{}]{}Proga, D., Stone, J. M., \& Drew, J. E. 1998, MNRAS, 295, 595
\bibitem[\protect\citeauthoryear{}{}]{}Proga, D., \& Kallman, T. R. 2002, ApJ, 565, 455
\bibitem[\protect\citeauthoryear{}{}]{}Proga, D., Kallman, T. R., Drew, J. E., \& Hartley, L. E. 2002, ApJ, 572, 382
\bibitem[\protect\citeauthoryear{}{}]{}Proga, D. 2007, ApJ, 661, 693
\bibitem[\protect\citeauthoryear{}{}]{}Proga, D., \& Kurosawa, R. 2009, AIPC, 1171, 295
\bibitem[\protect\citeauthoryear{}{}]{}P\^aris, I., Petitjean, P., Aubourg, \'E., et al. 2012, A\&A, 548, 66
\bibitem[\protect\citeauthoryear{}{}]{}Qin, Y. P., Chen, Z. F., L\"u, L. Z., et al. 2013, PASJ, 65, 8
\bibitem[\protect\citeauthoryear{}{}]{}Rodr\'Iguez Hidalgo, P., Hamann, F., \& Hall, P. 2011, MNRAS, 411, 247
\bibitem[\protect\citeauthoryear{}{}]{}Ross, N. P., Myers, A. D., Sheldon, E. S. et al., 2012, ApJS, 199, 3
\bibitem[\protect\citeauthoryear{}{}]{}Savage, B. D. \& Sembach, K. R. 1991, ApJ, 379, 245
\bibitem[\protect\citeauthoryear{}{}]{}Schneider, D. P., Richards, G. T., Hall, P. B. et al., 2010, AJ, 139, 2360
\bibitem[\protect\citeauthoryear{}{}]{}Shen, Y., Richards, G., Strauss, M. A., et al. 2011, ApJS, 194, 45
\bibitem[\protect\citeauthoryear{}{}]{}Tombesi, F., Cappi, M., Reeves, J. N., et al. 2010, A\&A, 521, A57
\bibitem[\protect\citeauthoryear{}{}]{}Tombesi, F., Cappi, M., Reeves, J. N., et al. 2011, ApJ, 742, 44
\bibitem[\protect\citeauthoryear{}{}]{}Verner, D. A., Barthel, . D., \& Tytler, D. 1994, A\&AS, 108, 287
\bibitem[\protect\citeauthoryear{}{}]{}Vivek, M., Srianand, R., Mahabal, A., \& Kuriakose, V. C. 2012, MNRAS, 421, L107
\bibitem[\protect\citeauthoryear{}{}]{}Weymann, R. J., Morris, S. L., Foltz, C. B., \& Hewett, P. C. 1991, ApJ, 373, 23
\bibitem[\protect\citeauthoryear{}{}]{}Wise, J. H., Eracleous, M., Charlton, J. C., \& Ganguly, R. 2004, ApJ, 613, 129
\bibitem[\protect\citeauthoryear{}{}]{}Wild, V., \& Hewett, P. C. 2005, MNRAS, 358, 1083
\bibitem[\protect\citeauthoryear{}{}]{}Wu, J., Charlton, J. C., Misawa, T., Eracleous, M., \& Ganguly, R. 2010, ApJ, 722, 997
\bibitem[\protect\citeauthoryear{}{}]{}York, D. G., Adelman, J., Anderson, J. E., Jr., et al. 2000, AJ, 120, 1579
\end{thebibliography}
\end{document}